\documentclass{birkmult}

\usepackage{graphicx}
\usepackage{dcolumn}
\usepackage{bm}
\usepackage{dsfont}
\usepackage{amsthm}
\usepackage{graphicx}

\theoremstyle{definition}
\newtheorem{Def}{Definition}[section]

\theoremstyle{plain}

\newtheorem{Thm}[Def]{Theorem}

\theoremstyle{remark}

\newtheorem*{Thm*}{}

\newcommand{\dd}{\mathrm{d}}
\newcommand{\ee}{\mathrm{e}}
\newcommand{\ii}{\mathrm{i}}
\newcommand{\rh}{r_\mathrm{h}}
\newcommand{\Hp}{{\mathcal H^+}}
\newcommand{\Hm}{{\mathcal H^-}}
\newcommand{\Hpm}{{\mathcal H^\pm}}
\newcommand{\E}{\mathcal E}
\newcommand{\f}{{\ee^{2U}}}

\newcommand{\C}{\mathds C}

\newcommand{\rW}{\varrho}
\newcommand{\zW}{\zeta}
\newcommand{\pmN}{^\pm_\mathrm{N}}
\newcommand{\pN}{^+_\mathrm{N}}
\newcommand{\mN}{^-_\mathrm{N}}
\newcommand{\pmS}{^\pm_\mathrm{S}}
\newcommand{\pS}{^+_\mathrm{S}}
\newcommand{\mS}{^-_\mathrm{S}}

\newcommand{\beq}{\begin{equation}}
\newcommand{\eeq}{\end{equation}}
\newcommand{\bea}{\begin{eqnarray}}
\newcommand{\eea}{\end{eqnarray}}

\begin{document}


\title[The inner Cauchy horizon of axisymmetric and stationary black holes]
{The inner Cauchy horizon of axisymmetric and
 stationary black holes with surrounding matter in Einstein-Maxwell
 theory: study in terms of soliton methods} 

\author{J\"org Hennig}
\address{Max Planck Institute for Gravitational Physics\\
Am M\"uhlenberg 1\\
D-14476 Golm\\
Germany}
\email{pjh@aei.mpg.de}
\author{Marcus Ansorg}
\address{Max Planck Institute for Gravitational Physics\\
Am M\"uhlenberg 1\\
D-14476 Golm\\
Germany}
\email{mans@aei.mpg.de}

\date{\today}

\begin{abstract}
 We use soliton methods in order to investigate the interior
 electrovacuum region of axisymmetric and 
 stationary, electrically charged black holes with arbitrary surrounding
 matter in Einstein-Maxwell theory. 
 These methods can be applied since the  
 Einstein-Maxwell vacuum equations permit the formulation in terms of the
 integrability condition of an associated linear matrix problem.
 We find that there always exists a regular inner Cauchy horizon inside the
 black hole, provided the angular momentum $J$ and charge $Q$
 of the black hole do not vanish simultaneously. Moreover, the soliton
 methods provide us with an explicit relation for the metric on the
 inner Cauchy horizon in terms of that on the event horizon. 
 In addition, our analysis reveals the remarkable universal relation
 $(8\pi J)^2+(4\pi Q^2)^2=A^+ A^-$, where 
 $A^+$ and $A^-$ denote the areas of event and inner Cauchy horizon respectively. 
\end{abstract}


\maketitle
\section{Introduction \label{sec:Intro}}

The single rotating, electrically charged,
axisymmetric and stationary Kerr-New\-man black hole in electrovacuum
is characterised by the existence of two different so-called Cauchy horizons $\mathcal H^\pm$.
One of these horizons is the well-known event horizon $\mathcal H^+$ which can be considered 
as a boundary of the exterior electrovacuum world. Outside the event horizon, the
Einstein-Maxwell equations take an elliptic form which is related to the fact that in this regime the two Killing vectors $\eta^i$ and $\xi^i$, describing axisymmetry and stationarity respectively, can be combined linearly to form a timelike vector, i.e.\footnote{In the formulae (\ref{eta_xi_lt_0}) and (\ref{eta_xi_gt_0}) and the corresponding discussion in the text, we exclude points located on the symmetry axis (the `rotation axis'). Note that the Killing vector $\eta^i$ vanishes identically on this axis which implies $\eta_{[i}\xi_{j]}\eta^i \xi^j = 0$ there.}
\begin{equation}\label{eta_xi_lt_0}
\eta_{[i}\xi_{j]}\eta^i \xi^j < 0.
\end{equation}
In contrast, on the event horizon any linear combination of the two Killing vectors is either space-like or null, 
\begin{equation}\label{eta_xi_eq_0}
\eta_{[i}\xi_{j]}\eta^i \xi^j = 0,
\end{equation}
i.e.~the horizon is a so-called Killing horizon:
\begin{equation}\label{KillongHor}
\chi^i\chi_i=0, \quad\chi^i=\xi^i+\omega^+\eta^i,
\end{equation}
where $\omega^+$ denotes the constant angular velocity of the black
hole's event horizon. This Killing horizon condition leads to specific
boundary conditions valid on the event horizon. While in this manner a
well-defined elliptic boundary problem of the Einstein-Maxwell equations
emerges, it is possible to extend its solution beyond the event horizon
into the electrovacuum interior of the black hole. Entering this region,
one recognizes that now for the two Killing vectors  
\begin{equation}\label{eta_xi_gt_0}
\eta_{[i}\xi_{j]}\eta^i \xi^j > 0
\end{equation}
holds, meaning that any non-trivial linear combination of the two Killing vectors is space-like. As a consequence, the Einstein-Maxwell vacuum  equations are hyberbolic in an inner vicinity of $\mathcal H^+$. Taking the boundary values on $\mathcal H^+$ as `initial data' for this hyperbolic system, one can `evolve' the vacuum solution regularly further into the black hole's interior. In this manner one finds, for the Kerr-Newman black holes, a `future boundary' of this hyperbolic region, that is the future boundary of the domain of dependence of the event horizon, i.e.~the inner Cauchy horizon $\mathcal H^-$.

Remarkably, the two horizons of the Kerr-Newman black holes exhibit an interesting relation, which becomes apparent through the equality
\begin{equation}\label{ApAm}
 (8\pi J)^2+(4\pi Q^2)^2 = A^+ A^-,
\end{equation}
where $J$ and $Q$ are angular momentum and charge of the black hole and $A^\pm$ denote the surface areas of the horizons $\mathcal H^\pm$. Note that for the Kerr-Newman black holes $\mathcal H^-$ is regular if and only if the left hand side of the above formula is strictly positive, i.e.~if $J$ and $Q$ do not vanish simultaneously. Then the black hole singularity is located further inside,
that is inside $\mathcal H^-$. In the limit $J\to 0, Q\to 0$ the singularity approaches the inner Cauchy horizon, i.e.~$\mathcal H^-$ becomes singular in this limit.

In pure Einsteinian gravity (i.e.~without Maxwell field), 
these observations have been generalized in \cite{Ansorg2}. It
was shown that for axisymmetric and stationary black holes {\em with arbitrary surrounding matter}
there exists a regular inner Cauchy horizon if and only if $J\neq 0$
holds. Moreover it was possible to identify a general relation between the two horizons $\mathcal H^\pm$ through which the metric on $\mathcal H^-$ is expressed explicitly in terms of that on $\mathcal H^+$. As a consequence of this explicit formula it turned out that all such black holes satisfy relation~\eqref{ApAm} (with $Q=0$).

It is the aim of this paper to carry this result over to the situation in which electromagnetic fields are included, i.e.~ to show that for axisymmetric and stationary, electrically charged black holes with arbitrary surrounding matter in Einstein-Maxwell theory
\begin{enumerate}
\item \label{statement_1}
	there exists a regular inner Cauchy horizon $\mathcal H^-$ if and only if angular momentum $J$ and charge $Q$
 of the black hole do not vanish simultaneously,
\item \label{statement_2}
	there is an explicit relation between the metric and electromagnetic quantities on the two horizons $\mathcal H^\pm$,
\item \label{statement_3}
	the universal formula~\eqref{ApAm} is valid.
\end{enumerate}
Thus, this paper provides a detailed description of the work presented in \cite{Hennig4}.

For the derivation of the pure Einsteinian results in \cite{Ansorg2} a particular soliton method was used -- the so-called 
B\"acklund transformation. It was possible to apply this method because the axisymmetric and stationary Einstein vacuum equations can be written in terms of the integrability condition of an associated linear matrix problem. The B\"acklund transformation utilizes this structure and creates a new solution from a previously known one. In \cite{Ansorg2} this procedure was the essential ingredient in writing an arbitrary regular axisymmetric, stationary black hole solution in terms of another solution, which describes a spacetime without a black hole, but with a completely regular central vacuum region. As a consequence of the symmetries of this regular solution, the desired relation between the two horizons was found.

Proceeding to the Einstein-Maxwell fields, we find that the applicability of the B\"acklund method seems limited. In particular, it is not straightforward to create in this manner the Kerr-Newman solutions from the flat Minkowski space, see \cite{Neugebauer}. Consequently, in this paper we treat the combined Einstein-Maxwell situation in a different way.

A feature common to both the pure Einstein and the combined Einstein-Maxwell cases is the existence, already mentioned, of an associated linear matrix problem whose integrability condition is equivalent to the field equations in vacuum, see  \cite{Neugebauer}. The B\"acklund transformation is merely one of several solution techniques (another one is the so-called `inverse scattering method', see \cite{Neugebauer2}) whose applicability results from the existence of this linear problem (LP). As will be described below, in the full Einstein-Maxwell situation the integration of the LP along the boundaries of the inner hyperbolic region yields sufficient information to derive the above statements \ref{statement_1}--\ref{statement_3}.

The paper is organized as follows. In Sec.~\ref{sec:Coords}, we introduce appropriate coordinates which are adapted to the subsequent analysis. We write the Einstein-Maxwell equations in terms of the Ernst formulation \cite{Ernst} for which the LP can be introduced. Moreover we list necessary horizon boundary and axis regularity conditions. In Sec.~\ref{sec:LP}, we describe the LP and, moreover, show that a similar LP can be found in a rotating frame of reference. The relation of the solution of the LP in the original to that in the rotating frame is derived explicitly. Then, in Sec.~\ref{solution}, we
determine the solution of the LP along the boundaries of the inner hyperbolic region, including the two horizons $\mathcal H^\pm$. For this treatment, the known event horizon boundary conditions are taken into account. The derivation of corresponding formulae in the two rotating frames of reference completes the analysis. In this way, an explicit formula relating metric and elctromagnetic potentials on $\mathcal H^-$ to those on $\mathcal H^+$ arises, see Sec.~\ref{sec:EP}.  As a further consequence of our study of the LP, we show in Sec.~\ref{sec:eq} the validity of Eq.~\eqref{ApAm} for axisymmetric and stationary, electrically charged black holes with arbitrary surrounding matter in Einstein-Maxwell theory. Finally, in Sec.~\ref{sec:Dis}, we conclude with a discussion.

\section{Coordinate systems and Einstein-Maxwell equations
\label{sec:Coords}}
\subsection{Weyl coordinates and Boyer-Lindquist-type coordinates}

We consider axisymmetric and stationary spacetimes, consisting of an electrically charged
central black hole and surrounding matter in Einstein-Maxwell theory. The
immediate vicinity of the black hole event horizon must be
electrovacuum, see \cite{Carter} and \cite{Bardeen}. In the following, we investigate 
the metric and electromagnetic potentials in such an electrovacuum region 
both inside and outside the black hole.

In the exterior electrovacuum vicinity we introduce Weyl coordinates\linebreak
$(\varrho, \zeta, \varphi, t)$ 
in which the line element reads as follows:
\begin{equation}\label{LE1}
 \dd s^2 = \ee^{-2U}\left[\ee^{2k}(\dd\rW^2+\dd\zW^2)
            +\rW^2\dd\varphi^2\right]
           -\ee^{2U}(\dd t+a\dd\varphi)^2.
\end{equation}

\begin{figure}
 \centering
 \includegraphics[scale=0.85]{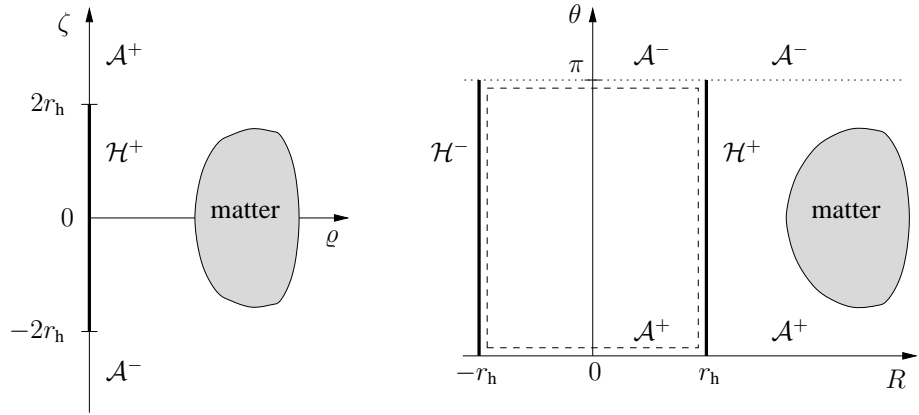}
 \caption{      Sketch of a part of a black hole spacetime
		in Weyl coordinates (\emph{left panel})
		and Boyer-Lindquist type coordinates (\emph{right
                panel}). $\mathcal 
		A^+$ and $\mathcal A^-$ denote upper and lower parts of the
		symmetry axis and $\mathcal H^+$ and $\mathcal H^-$
                denote event 
		and inner Cauchy horizons. In Sec.~\ref{solution} we will
                integrate the linear 
		problem for the Einstein-Maxwell equations along the
                closed dashed line (figure taken from \cite{Hennig4}). 
	}
 \label{fig1}
\end{figure}

The metric potentials $U$, $k$, and $a$ are functions of $\varrho$
and $\zeta$ alone. As sketched in Fig.~\ref{fig1} (left panel),
the event horizon $\Hp$ is located on the interval $-2\rh\le\zeta\le 2\rh$,
$\rh=\textrm{constant}$, of the $\zeta$-axis. The remaining part
$|\zeta|>2\rh$ of the $\zeta$-axis corresponds to the rotation
axis. In particular, we denote with $\mathcal A^+$ and $\mathcal A^-$ 
the axis sections where $\zeta\geq 2\rh$ and $\zeta\leq - 2\rh$ respectively.

The form \eqref{LE1} of the line element does not characterize uniquely a specific coordinate system. More precisely, if in our `original' system, denoted by $\Sigma$, the metric reads as in \eqref{LE1}, then in any frame $\Sigma'$, that rotates at a constant angular velocity $\omega_0$ with respect to $\Sigma$, the line element will assume the same structure. Note that in $\Sigma'$ the coordinates read $(\varrho, \zeta, \varphi', t)$, with the only new coordinate given by
\begin{equation}\label{phi_prime}
 \varphi' = \varphi-\omega_0 t.
\end{equation}
We will make use of this freedom and choose appropriate coordinate
systems $\Sigma$ and $\Sigma'$ in order to achieve the results of this
paper, that is the statements \ref{statement_1}--\ref{statement_3} in
Sec.~\ref{sec:Intro}. In particular, we will place ourselves in such an
original system $\Sigma$ in which both the event and inner Cauchy
horizon angular velocities $\omega^\pm$ do not vanish. More details on
this choice are presented in Sec.~\ref{rot}. 

In order to investigate the \emph{interior} of the black hole, which
is characterized by negative values of $\varrho^2$, we also introduce
Boyer-Lindquist-type coordinates $(R, \theta, \varphi, t)$ via
\begin{equation}\label{BoyerLcoord}
   \varrho^2 = 4(R^2-\rh^2)\sin^2\!\theta,\qquad
   \zeta  = 2R\cos\theta.
\end{equation}
(Note that, in the case of the Kerr-Newman black hole, these coordinates
are closely related to Boyer-Lindquist coordinates $(r,\theta,\varphi, t)$, where 
the only different coordinate is $r=2R+M$ with $M$ denoting the ADM mass of the spacetime.)

In the coordinates $(R, \theta, \varphi, t)$, the event horizon $\Hp$ is located at $R=\rh$. As we shall see below, the inner Cauchy
horizon $\Hm$ is characterized through $R=-\rh$, see Fig.~\ref{fig1} (right panel). It is the aim of this paper to show that both metric and electromagnetic quantities are regular in terms of $R$ and $\cos\theta$ within the interior vacuum region described by $-\rh\leq R\leq\rh$ (including $\Hm$), provided that $J$ and $Q$ do not both vanish.

For convenience, we introduce new metric functions that are, for a
regular black hole, analytic in terms of $R$ and $\cos\theta$ in the
black hole vicinity, see \cite{Bardeen},
\begin{subequations}\label{func}
\begin{eqnarray}
 \hat\mu & := & 4\ee^{2k-2U}(R^2-\rh^2\cos^2\!\theta),\\
 \hat u & := & 4(R^2-\rh^2)\ee^{-2U}-\frac{a^2}{\sin^2\!\theta}\ee^{2U},\\
 \omega & := & \frac{a\ee^{4U}}{4(R^2-\rh^2)\sin^2\!\theta-a^2\ee^{4U}}.
\end{eqnarray}
\end{subequations}
Moreover, $\hat\mu$ and $\hat u$ are strictly positive in that regime.
In terms of these functions, the Boyer-Lindquist-type line element is
given by
\begin{eqnarray}\label{LE2}
 \dd s^2  =  \hat\mu\left(\frac{\dd R^2}{R^2-\rh^2}+\dd\theta^2\right)
           +\hat u\sin^2\!\theta\,(\dd\varphi-\omega\dd t)^2
          -\frac{4}{\hat u}(R^2-\rh^2)\dd t^2.
\end{eqnarray}
\subsection{The Einstein-Maxwell equations}

In the electrovacuum region, the electromagnetic field alone constitutes the energy momentum tensor
\begin{equation}
 T_{ij} = \frac{1}{4\pi}\left(F_{ki}F^k_{\ j}
          -\frac{1}{4}g_{ij}F_{kl}F^{kl}\right),
\end{equation}
where $F_{ij}$ is the electromagnetic field tensor.
We use the Lorenz gauge, in which $F_{ij}$ can be written in terms of a
vector potential $(A_i) = (0, 0, A_{\varphi}, A_t)$,
\begin{equation}
 F_{ij}=A_{i,j}-A_{j,i}.
\end{equation}
Note that, like the metric quantities, $A_\varphi$ and $A_t$ also depend on $R$
and $\theta$ only.

We introduce the complex electromagnetic potential $\Phi$ and the complex
Ernst potential $\E$ \cite{Ernst, ExactSolutions} by 
\begin{equation}\label{cpot}
 \Phi = A_t + \ii\beta,\quad
 \E   =\ee^{2U}-|\Phi|^2+\ii b,
\end{equation}
where the imaginary parts $b$ and $\beta$ are related to metric and
vector potentials via
\begin{subequations}\label{ab1}
\begin{eqnarray}
 \label{a1}
 a_{,\varrho}  & = & \varrho\ee^{-4U}\left[b_{,\zeta}
                  -\ii(\bar\Phi\Phi_{,\zeta}-\Phi\bar\Phi_{,\zeta})\right],\\
 \label{a2}
 a_{,\zeta} & = & -\varrho\ee^{-4U}\left[b_{,\varrho}
                  -\ii(\bar\Phi\Phi_{,\varrho}-\Phi\bar\Phi_{,\varrho})\right],\\
 \label{b1}
 \beta_{,\varrho} & = &
 \frac{\ee^{2U}}{\varrho}(aA_{t,\zeta}-A_{\varphi,\zeta}),\\
 \label{b2}
  \beta_{,\zeta} & = &
 -\frac{\ee^{2U}}{\varrho}(aA_{t,\varrho}-A_{\varphi,\varrho}),
\end{eqnarray}
\end{subequations}
or, in terms of $R$ and $\theta$,
\begin{subequations}\label{ab2}
\begin{eqnarray}
 \label{a3}
 a_{,R} & = & -2\sin\theta\,\ee^{-4U}\left[b_{,\theta}
            -\ii(\bar\Phi\Phi_{,\theta}-\Phi\bar\Phi_{,\theta})\right],\\
 \label{a4}
 a_{,\theta} & = & 2(R^2-\rh^2)\sin\theta\,\ee^{-4U}
           \left[b_{,R}
          -\ii(\bar\Phi\Phi_{,R}-\Phi\bar\Phi_{,R})\right],\\
 \label{b3}
 \beta_{,R} & = & -\frac{\ee^{2U}}{2(R^2-\rh^2)\sin\theta}
                  (aA_{t,\theta}-A_{\varphi,\theta}),\\
 \label{b4}
 \beta_{,\theta} & = & \frac{\ee^{2U}}{2\sin\theta}
                  (aA_{t,R}-A_{\varphi,R}).                 
\end{eqnarray}
\end{subequations}

In this formulation, the Einstein-Maxwell equations in electrovacuum are
equivalent to the two complex Ernst equations \cite{Ernst}
\begin{subequations}\label{EM}
\begin{eqnarray}
  \f\bigtriangleup\E & = &  \nabla\E\cdot(\nabla\E+2\bar\Phi\nabla\Phi),
  \\[1.5ex]
  \f\bigtriangleup\Phi & = & \nabla\Phi\cdot(\nabla\E+2\bar\Phi\nabla\Phi).
\end{eqnarray}
\end{subequations}
Here, $\bigtriangleup$ and $\nabla$ denote Laplace and nabla
operators in flat cylindrical coordinates $(\varrho,\zeta,\varphi)$. In
terms of $R$ and $\theta$, these equations take the form
\begin{subequations}\label{EM_2}
\begin{align}
 & \f\left[(R^2-\rh^2)\E_{,RR}+2R\E_{,R}+\E_{,\theta\theta}
  +\cot\theta\,\E_{,\theta}\right]\nonumber\\
 & \quad = (R^2-\rh^2)(\E_{,R}+2\bar\Phi\Phi_{,R})\E_{,R}
  +(\E_{,\theta}+2\bar\Phi\Phi_{,\theta})\E_{,\theta}\ ,\\[1.5ex]
 & \f\left[(R^2-\rh^2)\Phi_{,RR}+2R\Phi_{,R}+\Phi_{,\theta\theta}
  +\cot\theta\,\Phi_{,\theta}\right]\nonumber\\
 &  \quad = (R^2-\rh^2)(\E_{,R}+2\bar\Phi\Phi_{,R})\Phi_{,R}
   +(\E_{,\theta}+2\bar\Phi\Phi_{,\theta})\Phi_{,\theta}\ .
\end{align}
\end{subequations}
Note that these equations are elliptic for $|R|>\rh$ but degenerate
at $R=\pm\,\rh$. Only in the interior region $-\rh<R<\rh$ are these equations hyperbolic, i.e.~in these 
coordinates the inner Cauchy horizon $\Hm, R=-\rh$ is a `future boundary' of this hyperbolic vacuum region, that is the future boundary of the domain of dependence of the event horizon $\Hp, R=+\,\rh$, see Fig.~\ref{fig1} (right panel).

\subsection{Boundary and regularity conditions\label{sec:BC}}

In this section we summarize particular horizon boundary and axis regularity conditions, which are essential in the forthcoming analysis. At $\mathcal H^\pm$ the following conditions are satisfied (cf. \cite{Bardeen}):
\begin{subequations}\label{BC}
\begin{equation}
 \omega=-\frac{1}{a}=\textrm{constant}=\omega^\pm\neq 0,
\end{equation}
\begin{equation}
 \frac{ 2\rh}{\sqrt{\hat\mu\hat u}} = \textrm{constant}=\pm\kappa^\pm>0,
\end{equation}
\begin{equation}
  \phi=\textrm{constant}=\phi^\pm.
\end{equation}
\end{subequations}
Here $\omega^\pm$ and $\kappa^\pm$ denote the constant horizon angular velocities and horizon
surface gravities respectively, and the {\em comoving} electric potential $\phi$ is defined by
\begin{equation}\label{phi}
 \phi = A_t + \omega A_\varphi.
\end{equation}

As mentioned already, we choose a coordinate frame $\Sigma$ in which both horizon angular velocities $\omega^\pm$ do not vanish (see discussion in Sec.~\ref{rot}).

The surface gravities are required to be different from zero, since, in this paper, we exclude degenerate black holes for which $\mathcal H^-$ and $\mathcal H^+$ coincide and the hyperbolic region disappears, i.e.~we assume $\rh>0$. 

`North pole' and `south pole' of the two horizons $\mathcal H^\pm$ are characterized by $R=\pm\,\rh,\theta=0$ and $R=\pm\rh,\theta=\pi$ respectively. At these points, the horizons meet the rotational axis and
the following regularity conditions hold:
\begin{equation}\label{RC}
 \hat\mu\pmN = \hat\mu\pmS = \hat u\pmN = \hat u\pmS
 = \pm\frac{2\rh}{\kappa^\pm},\quad
 \ee^{2U}=0.
\end{equation}
The notation $(\cdot)_\mathrm{N}$ and $(\cdot)_\mathrm{S}$ discriminates between the
values at north and south pole.

In addition to these conditions, on the portions of the axis $\mathcal A^\pm$, we have
\begin{equation}\label{RC1}
 A_\varphi=0,\quad
 a=0,\quad \E_{,\theta}=0,\quad \Phi_{,\theta}=0.
\end{equation}

\section{The linear problem\label{sec:LP}}

The Ernst equations \eqref{EM} belong to a remarkable class of
physically relevant nonlinear partial differential equations, which are characterized by the existence of an associated linear problem (LP) whose integrability conditions are
equivalent to the differential equation in question\footnote{Other
examples of such equations are the Korteweg-de Vries equation, the
sine-Gordon equation and the nonlinear Schr\"{o}dinger equation.}. A careful study of this LP will provide us with the
information needed to derive the statements
\ref{statement_1}--\ref{statement_3} listed in Sec. \ref{sec:Intro}. 
 
For the formulation of the LP, which is associated with the Ernst equations \eqref{EM},
we introduce the complex coordinates
\begin{equation}
 z=\varrho + \ii\zeta,\quad \bar z = \varrho - \ii\zeta
\end{equation}
as well as the function
\begin{equation}\label{la}
 \lambda(K,z,\bar z)=\sqrt{\frac{K-\ii\bar z}{K+\ii z}},
\end{equation}
which depends on the \emph{spectral parameter} $K\in\C$.
For fixed values $z$, $\bar z$, equation \eqref{la} describes a spectral
mapping $\C\to\C$,
$K\mapsto\lambda$ from a two-sheeted Riemann surface ($K$-plane)
onto the complex $\lambda$-plane. `Upper' and `lower' $K$-sheets 
(defined by $\lambda=\pm 1$ for $K=\infty$) are connected
at the two branch points $K_1=\ii\bar z$ ($\lambda=0$)
and $K_2=-\ii z$ ($\lambda=\infty$).

The LP is a system of first order differential equations for a $3\times
3$ matrix pseudopotential ${\bf\Omega}={\bf\Omega}(K,z,\bar z)$, which
reads \cite{Neugebauer}
\begin{equation}\label{LP}
 {\bf\Omega}_{,z} = {\bf M}{\bf \Omega},\qquad
 {\bf\Omega}_{,\bar z} = {\bf N}{\bf\Omega}, 
\end{equation}
where
\begin{eqnarray*}
 {\bf M} & = & \left(\begin{array}{ccc}
                  B_1  & 0   & E_1\\
                  0    & A_1 & 0  \\
                  -F_1 & 0   & \frac12(A_1 + B_1)
                  \end{array}\right)
               + \lambda\left(\begin{array}{ccc}
                  0    & B_1 & 0\\
                  A_1  & 0   & -E_1\\
                  0    & -F_1& 0
                  \end{array}\right),\\[1.5ex]
 {\bf N} & = & \left(\begin{array}{ccc}
                  B_2  & 0   & E_2\\
                  0    & A_2 & 0  \\
                  -F_2 & 0   & \frac12(A_2 + B_2)
                  \end{array}\right)
               + \frac{1}{\lambda}\left(\begin{array}{ccc}
                  0    & B_2 & 0\\
                  A_2  & 0   & -E_2\\
                  0    & -F_2& 0
                  \end{array}\right).             
\end{eqnarray*}
The matrix elements of ${\bf M}$ and ${\bf N}$ are functions of $z$ and
$\bar z$. In terms of the potentials $\E$ and $\Phi$ they are given by
\begin{subequations}
\begin{eqnarray}
 & A_1 =  \frac{1}{2}\ee^{-2U}(\E_{,z}+2\bar\Phi\Phi_{,z}),\quad
 E_1 = \ii\ee^{-U}\Phi_{,z},\\[1.5ex]
 & B_1 =  \frac{1}{2}\ee^{-2U}(\bar\E_{,z}+2\Phi\bar\Phi_{,z}),\quad
 F_1 = \ii\ee^{-U}\bar\Phi_{,z},\\[1.5ex]
 & A_2 =  \frac{1}{2}\ee^{-2U}(\E_{,\bar z}+2\bar\Phi\Phi_{,\bar z}),\quad
 E_2 = \ii\ee^{-U}\Phi_{,\bar z},\\[1.5ex]
 & B_2 =  \frac{1}{2}\ee^{-2U}(\bar\E_{,\bar z}+2\Phi\bar\Phi_{,\bar z}),\quad
 F_2 = \ii\ee^{-U}\bar\Phi_{,\bar z},
\end{eqnarray}
\end{subequations}
where $\f$ can be calculated from $\E$ and $\Phi$,
\[ \f = \Re\E+|\Phi|^2. \]
From the integrability condition
\[{\bf\Omega}_{,z\bar z}={\bf\Omega}_{,\bar z z}\] 
of the LP \eqref{LP} one derives equations that are 
equivalent to the Ernst equations \eqref{EM}. 

The pseudopotential ${\bf\Omega}$ is not uniquely determined by \eqref{LP}.
If ${\bf\Omega}$ is a solution, then ${\bf\Omega
C}(K)$ is also a solution for every $3\times 3$ matrix function ${\bf C}(K)$.
We can always find a ${\bf C}(K)$ to bring ${\bf\Omega}$ into the form
\begin{subequations}\label{gauge}
\begin{equation}
 {\bf\Omega}^>(K, z,\bar z)=\left(\begin{array}{crr}
              \psi_1^>(K,z,\bar z) &  \psi_1^<(K, z, \bar z) & 0\\
              \psi_2^>(K,z, \bar z) & -\psi_2^<(K, z, \bar z) & 0\\
              \psi_3^>(K,z, \bar z) &  \psi_3^<(K,z,\bar z) & 0
              \end{array}\right)
\end{equation}
and
\begin{equation}
 {\bf\Omega}^<(K, z,\bar z)=\left(\begin{array}{crr}
              \psi_1^<(K,z,\bar z)  &  \psi_1^>(K, z, \bar z) & 0\\
              \psi_2^<(K,z, \bar z) & -\psi_2^>(K, z, \bar z) & 0\\
              \psi_3^<(K,z, \bar z) &  \psi_3^>(K,z,\bar z) & 0
              \end{array}\right)\hspace{-0.8mm},
\end{equation}
\end{subequations}
which depends on three functions $\psi_1$, $\psi_2$, $\psi_3$. Here,
the superscript `$>$' or `$<$' indicates whether the functions
are evaluated in the upper ($\lambda=1$ for $K=\infty$) or
lower ($\lambda=-1$ for $K=\infty$) sheet of the two-sheeted Riemann $K$-surface.
Obviously, ${\bf\Omega}$ of this form is not
invertible. Nevertheless, we will see that it still contains sufficient
information about $\E$ and $\Phi$.
\subsection{Rotating frames of reference\label{rot}}

It turns out that the analysis of the LP, performed alone in the coordinate system $\Sigma$ with coordinates $(\varrho, \zeta, \varphi, t)$, does not give sufficient information about the relation of the potentials at the two horizons. However, the missing information can be obtained by studying the situation in the two frames $\Sigma^\pm$ of reference which rotate with the constant horizon angular velocities $\omega^\pm$ (cf.~\eqref{BC}) with respect to $\Sigma$. Since for the complete investigation $\Sigma$ needs to be different from $\Sigma^\pm$, we assume $\omega^\pm\neq 0$, a choice that can always be made because of the freedom with respect to our original system $\Sigma$, see discussion in Sec.~\ref{sec:Coords}. Hence, in the formulae appearing below we can safely divide by $\omega^\pm$.

As discussed in Sec.~\ref{sec:Coords}, in the coordinate system
$\Sigma'$ with coordinates $(\varrho, \zeta, \varphi', t)$   (see
\eqref{phi_prime}) and rotating at the constant angular velocity
$\omega_0$, the line element possesses again the structure
\eqref{LE1}. In particular, the corresponding potentials $U'$, $a'$, and
$k'$ are given in terms of $U$, $a$, and $k$ by 
\begin{subequations}\label{Ua}
\begin{eqnarray}
 \ee^{2U'} & = & \left[(1+\omega_0 a)^2
           -\omega_0^2\varrho^2\ee^{-4U}\right]\ee^{2U},\\[1.5ex]
  a'  & = &  \frac{a(1+\omega_0 a)-\omega_0\varrho^2\ee^{-4U}}
           {(1+\omega_0 a)^2-\omega_0^2\varrho^2\ee^{-4U}},\\[1.5ex]
 \ee^{2k'} & = & \left[(1+\omega_0 a)^2
           -\omega_0^2\varrho^2\ee^{-4U}\right]\ee^{2k}.
\end{eqnarray}
\end{subequations}
The components of the vector potential in the rotating system read
\begin{equation}
 A_\varphi' = A_\varphi,\quad A_t' = A_t+\omega_0 A_\varphi.
\end{equation}
As a consequence, the metric and electromagnetic potentials 
in $\Sigma'$ again satisfy the Ernst equations \eqref{EM} (in terms of corresponding potentials $\E'$ and $\Phi'$).
Hence an associated LP of the form \eqref{LP} can be found (with matrices ${\bf \Omega}', {\bf M}'$ and ${\bf N}'$).
From \eqref{ab1} and \eqref{Ua} it follows that the components of the matrices ${\bf M}'$,  ${\bf N}'$ read as
\begin{subequations}
\begin{align}
 & A'_1 = \frac{c_+ A_1-d}{c_-},\quad
   E'_1 = \sqrt{\frac{c_+}{c_-}}E_1,\\
 & B'_1 = \frac{c_- B_1+d}{c_+},\quad
   F'_1 = \sqrt{\frac{c_-}{c_+}}F_1,\\
 & A'_2 = \frac{c_- A_2+d}{c_+},\quad
   E'_2 = \sqrt{\frac{c_-}{c_+}}E_2,\\
 & B'_2 = \frac{c_+ B_2-d}{c_-},\quad
   F'_2 = \sqrt{\frac{c_+}{c_-}}F_2,
\end{align}
\end{subequations}
with
\begin{equation}\label{cpm}
 c_\pm := 1+\omega_0(a \pm \varrho\ee^{-2U}),\quad
 d := \frac{\omega_0}{2}\ee^{-2U}.
\end{equation}
The pseudopotential ${\bf\Omega}'$ in $\Sigma'$ arises as the solution of the LP \eqref{LP}, written in terms of 
${\bf M}'$ and ${\bf N}'$. It is, however, possible to establish a direct relation between ${\bf\Omega}'$ and ${\bf\Omega}$.
As an {\em ansatz}, we write
\begin{equation}\label{ansatz}
 {\bf\Omega}' = {\bf T\Omega}.
\end{equation}
where ${\bf T}$ is an unknown transformation matrix.
Combining \eqref{LP} and the corresponding equations for the
LP written in $\Sigma'$, we conclude that ${\bf T}$ has to obey the equations
\begin{subequations}
\begin{align}
 {\bf T}_{,z}+{\bf TM} - {\bf M}'{\bf T} = 0,\\
 {\bf T}_{,\bar z}+{\bf TN} - {\bf N}'{\bf T} = 0.
\end{align}
\end{subequations}
A solution, which yields via \eqref{ansatz} a pseudopotential ${\bf\Omega}'$ that possesses again the special structure \eqref{gauge}, turns out to be
\begin{eqnarray}\label{T}
 {\bf T} =  \left(\begin{array}{ccc}
          c_- & 0    & 0\\
          0   & c_+  & 0\\
          0   & 0    & \sqrt{c_+ c_-}
          \end{array}\right)
         +\ii(K+\ii z)\omega_0\ee^{-2U}
          \left(\begin{array}{ccc}
          -1      & -\lambda & 0\\
          \lambda & 1        & 0\\
          0       & 0        & 0
          \end{array}\right).
\end{eqnarray}
Note that this transformation matrix $\bf T$ is a generalization of a corresponding 
expression given in \cite{Neugebauer2000,Neugebauer2} in pure Einsteinian gravity (without Maxwell field).

\section{Solution of the linear problem\label{solution}}

As we derive in detail below, the relations of the metric and electromagnetic field quantities at the
inner Cauchy horizon $\mathcal H^-$ to those at the event horizon $\mathcal H^+$ 
emerge from the integration of the LP along the dashed lines in Fig.~\ref{fig1}
(right panel). This integration path contains the parts $(-\rh\le R\le\rh, \sin\theta=0)$ of the axis $\mathcal A^\pm$ as well as the two horizons $\mathcal H^\pm$. We are able to perfom this integration given that $\E$ and $\Phi$ are analytic with respect to $R$ and $\cos\theta$ in an {\em exterior} vicinity of $\mathcal H^+$ (including $\mathcal H^+$). Then, $\E$ and $\Phi$ can be expanded into an \emph{interior} vicinity of $\mathcal H^+$. Now, with regular
data on a slice inside the black hole, a theorem by Chru\'sciel (theorem 6.3 in \cite{Chrusciel1}\footnote{We obtain Chru\'sciels form of the line element by substituting $R=\rh\cos T$ and $\theta=\psi$.}) can be applied.
Although this theorem is formulated in pure Einsteinian gravity, the arguments presented in \cite{Chrusciel1} permit
a generalisation to the Einstein-Maxwell case considered here \cite{Chrusciel2}. The theorem assures that $\E$ and $\Phi$ exist and are regular for all values 
\[(R,\cos\theta)\in(-\rh,\rh]\times[-1,1],\]
i.e.~in the entire inner region between the horizons (see Fig.~\ref{fig1}), only excluding, for the time being, the inner Cauchy horizon $\mathcal H^-$. 

Along the entire integration path we have $\varrho=0$, cf.~\eqref{BoyerLcoord}. We study the LP for $\lambda=1$, that is in the upper sheet of the $K$-plane\footnote{With the gauge \eqref{gauge}, the solution of the LP in the lower sheet 
(in which $\lambda=-1$) can easily be obtained from that in the upper sheet.}.
Then, the LP reduces to an ODE with the general solution
\begin{equation}\label{sol}
 {\bf\Omega} = \left(\begin{array}{ccc}
               \bar\E+2|\Phi|^2 & 1  & \Phi\\
               \E               & -1 & -\Phi\\
               -2\ii\ee^U\bar\Phi & 0 & -\ii\ee^U
               \end{array}\right){\bf C}(K).
\end{equation}
Here, ${\bf C}$ is a $3\times 3$ matrix which depends on $K$ only. 
Respecting the gauge \eqref{gauge}, the third column of ${\bf C}$ vanishes.

It turns out that the regularity of the potentials $\E$ and $\Phi$ in $(R,\cos\theta)\in(-\rh,\rh]\times[-1,1]$ enables us perform the integration of the LP along $\mathcal A^\pm$ and $\mathcal H^+$. Moreover, a careful study of the LP for points on the integration path in the vicinities of the north and south poles of $\mathcal H^+$ reveals that the pseudopotentials possess specific continuity conditions there. In this way it becomes possible to derive the pseudopotentials on $\mathcal H^+$ and $\mathcal A^-$ in terms of expressions valid at $\mathcal A^+$. Proceeding now to $\mathcal H^-$ one finds that again the LP exhibits the explicit solution \eqref{sol} and, most importantly, permits a continuous link of this solution to the pseudopotentials at the two axes sections $\mathcal A^\pm$, which are joined to $\mathcal H^-$ at the inner Cauchy horizon's north and south poles. As a result, the regularity of the potentials $\E$ and $\Phi$ at $\mathcal H^-$ emerges, and their values can be found entirely in terms of those at $\mathcal H^+$. This procedure breaks down only if a specific parameter combination $B$ (see Eq.~\eqref{B_eqn} below) becomes infinite, which in turn happens if and only if both angular momentum $J$ and charge $Q$ vanish. 

We discuss the solutions of the LP on the various sections of the integration path and show that they can be expressed in terms of the three functions $C_1(K), C_2(K)$ and $C_3(K)$, which are introduced in Sec.~\ref{subsec:Ap} as `integration constants' of the LP. Moreover, specific north and south pole boundary values of the potentials as well as the constants $\omega^\pm(\neq 0)$ appear in the integration procedure. In the course of the investigation we find that these values satisfy specific relations. In particular, values at $\mathcal H^-$ can be written completely in terms of those at $\mathcal H^+$. It thus becomes possible to express the inner Cauchy horizon potentials entirely in terms of the event horizon potentials, see Sec.~\ref{sec:EP}.

\subsection{Solution on $\mathcal A^+$\label{subsec:Ap}}

As expressions valid on $\mathcal H^\pm$ become more concise if they are expressed in terms of those at an axis portion that joins the two horizons, we start our considerations of the solution of the LP on $\mathcal A^+$.

The gauge \eqref{gauge} does not completely fix the solution of the LP.
We obtain a unique solution by imposing the {\em normalization
conditions}\footnote{Note that the conditions \eqref{psi_j} are chosen
in accordance with regular solvability of the LP along the entire
integration path in a complex vicinity of the interval $[-2\rh,2\rh]$ of
the real $K$-axis. As in our analysis only values at
$K=\zW\in[-2\rh,2\rh]$ will be considered, such a vicinity is sufficient,
see Sec.~\ref{sec:EP} and, in particular, Eq.~\eqref{sheet}.}
\begin{equation}\label{psi_j}
\psi_1=\psi,\quad \psi_2=\psi,\quad \psi_3=0
\end{equation}
with
\begin{equation}
 \psi:=(K^2-4\rh^2)^3
\end{equation}
at some point $\zeta=\zeta_0$ on $\mathcal A^+$ 
in the lower sheet of the $K$-plane ($\lambda=-1$). 
As a consequence of \eqref{sol}, the normalization conditions
\eqref{psi_j} are then satisfied 
\emph{everywhere} on $\mathcal A^+$ in the lower sheet, and the solution
of the LP on $\mathcal A^+$ in the upper sheet (with $\lambda=1$) reads
\begin{equation}\label{sol1}
 {\bf\Omega} = \left(\begin{array}{ccc}
               \bar\E+2|\Phi|^2 & 1  & \Phi\\
               \E               & -1 & -\Phi\\
               -2\ii\ee^U\bar\Phi & 0 & -\ii\ee^U
               \end{array}\right)
               \left(\begin{array}{ccc}
               C_1(K) & 0 & 0\\
               C_2(K) & \psi(K) & 0\\
               C_3(K) & 0 & 0
               \end{array}\right).
\end{equation}
Here the three `integration constants' $C_1$, $C_2$,
$C_3$ (depending on $K$) appear.

Necessary additional information can be gathered by solving the LP for the corresponding
pseudopotentials in the two rotating frames of reference with $\omega_0=\omega^\pm\neq 0$ (see Sec.~\ref{rot}). With $\varrho=0$ and $a=0$ at $\mathcal A^+$ (cf.~\eqref{BC}) we obtain from \eqref{ansatz}, \eqref{T}, and \eqref{sol1}
\begin{eqnarray}\label{sol1a}
 {\bf\Omega}' = \left(\begin{array}{ccc}
               \bar\E+2|\Phi|^2-2\ii\omega^\pm(K-\zeta) & 1  & \Phi\\
               \E+2\ii\omega^\pm(K-\zeta)               & -1 & -\Phi\\
               -2\ii\ee^U\bar\Phi & 0 & -\ii\ee^U
               \end{array}\right)
              \left(\begin{array}{ccc}
               C_1(K) & 0 & 0\\
               C_2(K) & \psi(K) & 0\\
               C_3(K) & 0 & 0
               \end{array}\right).
\end{eqnarray}
\subsection{Solution on $\mathcal H^+$\label{subsec:Hp}}

On the event horizon, the solution of the LP yields in the upper sheet, i.e.~for $\lambda=1$:
\begin{equation}\label{sol2}
 {\bf\Omega} = \left(\begin{array}{ccc}
               \bar\E+2|\Phi|^2 & 1  & \Phi\\
               \E               & -1 & -\Phi\\
               -2\ii\ee^U\bar\Phi & 0 & -\ii\ee^U
               \end{array}\right)             
               \left(\begin{array}{ccc}
               D_1(K) & D_4(K) & 0\\
               D_2(K) & D_5(K) & 0\\
               D_3(K) & D_6(K) & 0
               \end{array}\right)
\end{equation}
with `integration constants' $D_1,\dots,D_6$.

In the rotating system with $\omega_0=\omega^+\neq 0$ we obtain (respecting $\varrho=0$ and $a=-1/\omega^+$ on $\mathcal H^+$)
\begin{equation}\label{sol2b}
 {\bf\Omega}' = 2\ii\omega^+(K-\zeta)
                \left(\begin{array}{ccc}
                -D_1(K) & -D_4(K) & 0\\
                D_1(K) & D_4(K) & 0\\
                0 & 0 & 0
                \end{array}\right),
\end{equation}
see \eqref{cpm}, \eqref{ansatz} and \eqref{T}.

We now derive six equations that provide us with $D_1,\ldots,D_6$ in terms of
$C_1, C_2, C_3$. These equations follow from the thorough discussion of
the LP for points on the integration path in the vicinities of the north
and south poles of $\mathcal H^+$. In particular, we find that both
${\bf\Omega}$ and ${\bf\Omega'}$ (i.e.~$\psi_1$, $\psi_2$, $\psi_3$ as
well as $\psi_1'$ $\psi_2'$, $\psi_3'$) are continuous there. One might
expect that the continuity of ${\bf\Omega}$ alone suffices for this
investigation, since ${\bf\Omega}$ has, in general, six non-trivial
components ($\psi_i^>$ and $\psi_i^<$, $i=1,2,3$, see \eqref{gauge}).  
However, since $\ee^{2U}=0$ at the north and south poles, the LP
degenerates there and as a consequence we obtain only two independent
equations. Hence, we have to supply this study with further
information. We obtain two further independent equations by
requiring the continuity of ${\bf\Omega}'$. But as another two equations
are needed in order to complete the analysis, we finally consider the
continuity of the two expressions $\ee^{-U}\psi^>_3$ and
$\ee^{-U}\psi^<_3$ at the north and south poles which again arises as a
consequence of the LP. In this manner we gather six independent
equations, which allow us to express $D_1,\ldots,D_6$ in terms of $C_1,
C_2, C_3$: 
\begin{subequations}\label{D}
\begin{eqnarray}
 D_1 & = & C_1 -\ii\frac{\E\pN C_1-C_2-\Phi\pN C_3}{2\omega^+(K-2\rh)},\\
 D_2 & = & C_2-\ii\frac{(\E\pN+2|\Phi\pN|^2)(\E\pN C_1-C_2-\Phi\pN C_3)}
                   {2\omega^+(K-2\rh)},\qquad\\
 D_3 & = & C_3+\ii\bar\Phi\pN\frac{\E\pN C_1-C_2-\Phi\pN C_3}
                              {\omega^+(K-2\rh)},\\
 D_4 & = & \frac{\ii\psi}{2\omega^+(K-2\rh)},\\
 D_5 & = & \psi\left(1+\ii\frac{\E\pN+2|\Phi\pN|^2}{2\omega^+(K-2\rh)}
           \right),\\
 D_6 & = & -\ii\psi\frac{\bar\Phi\pN}{\omega^+(K-2\rh)}.
\end{eqnarray}
\end{subequations}

\subsection{Solution on $\mathcal A^-$\label{subsec:Am}}

For the section $\mathcal A^-$ we write
\begin{equation}\label{sol4}
 {\bf\Omega} = \left(\begin{array}{ccc}
               \bar\E+2|\Phi|^2 & 1  & \Phi\\
               \E               & -1 & -\Phi\\
               -2\ii\ee^U\bar\Phi & 0 & -\ii\ee^U
               \end{array}\right)\tilde{\bf C}
\end{equation}
for the pseudopotential in the upper sheet ($\lambda=1$).
Here
\begin{equation}
 \tilde{\bf C}(K) = \left(\begin{array}{ccc}
               \tilde C_1(K) & \tilde C_4(K) & 0\\
               \tilde C_2(K) & \tilde C_5(K) & 0\\
               \tilde C_3(K) & \tilde C_6(K) & 0
               \end{array}\right).
\end{equation}
In the rotating system with $\omega_0=\omega^\pm\neq 0$ (cf.~Sec.~\ref{rot}) we have
\begin{equation}\label{sol4a}
 {\bf\Omega}' = \left(\begin{array}{ccc}
               \bar\E+2|\Phi|^2-2\ii\omega^\pm(K-\zeta) & 1  & \Phi\\
               \E+2\ii\omega^\pm(K-\zeta)               & -1 & -\Phi\\
               -2\ii\ee^U\bar\Phi & 0 & -\ii\ee^U
               \end{array}\right)
               \tilde{\bf C}(K).
\end{equation}
As continuity properties valid at the south pole of $\mathcal H^+$ follow from the LP again,
we are able to express the `integration constants' $\tilde C_1,\dots,\tilde C_6$.
At first, $\tilde C_1,\dots,\tilde C_6$ can be found in terms of $D_1,\dots,D_6$. Then, via \eqref{D}, the $\tilde C_1,\dots,\tilde C_6$ arise as functions of $C_1$, $C_2$, $C_3$.

\subsection{Solution on $\mathcal H^-$}\label{subsec:Hm}

The solution of the LP on $\mathcal H^-$ is again given by the general structure \eqref{sol}. Hence we may write
\begin{equation}\label{sol3}
 {\bf\Omega} = \left(\begin{array}{ccc}
               \bar\E+2|\Phi|^2 & 1  & \Phi\\
               \E               & -1 & -\Phi\\
               -2\ii\ee^U\bar\Phi & 0 & -\ii\ee^U
               \end{array}\right)
               \left(\begin{array}{ccc}
               \tilde D_1(K) & \tilde D_4(K) & 0\\
               \tilde D_2(K) & \tilde D_5(K) & 0\\
               \tilde D_3(K) & \tilde D_6(K) & 0
               \end{array}\right)
\end{equation}
for the pseudopotential in the upper sheet ($\lambda=1$) at the inner Cauchy horizon $\mathcal H^-$.
In the rotating system with $\omega_0=\omega^-\neq 0$ (cf.~Sec.~\ref{rot}) we have
\begin{equation}\label{sol3b}
 {\bf\Omega}' = 2\ii\omega^-(K-\zeta)
                \left(\begin{array}{ccc}
                -\tilde D_1(K) & -\tilde D_4(K) & 0\\
                \tilde D_1(K) & \tilde D_4(K) & 0\\
                0 & 0 & 0
                \end{array}\right).
\end{equation}
As we cannot assume from the outset that the pseudopotential is regular at $\mathcal H^-$ and in particular at its north pole, we carefully study whether the LP can be solved on the integration path in the vicinity of this point. We find that this is indeed the case and that, moreover, specific continuity properties can be fulfilled which hold at the pole. These properties are similar to those valid on the poles of $\mathcal H^+$, see discussion in Sec.~\ref{subsec:Hp}. As a consequence, the quantities $\tilde D_1,\dots,\tilde D_6$ can be derived in terms of $C_1$, $C_2$, $C_3$. Note that 
the expressions for $\tilde D_i$ are of the form \eqref{D}, with $\rh$ and the superscript `$+$' replaced by $-\rh$ and `$-$' respectively.

In a similar manner we may calculate the $\tilde D_i$ from continuity conditions studied at the south
pole of $\mathcal H^-$. Consequently, we obtain \emph{two} different systems for the $\tilde D_i$, and the requirement of equality of these two sets leads us to the following relations
\begin{subequations}\label{minus}
\begin{eqnarray}
 \omega^- & = & \omega^+(1-2AB),\\
 b\mN     & = & b\pN-\left[(b\pN-b\pS)A
                -(\beta\pN+\beta\pS)(\beta\pN-\beta\pS)^3\right]B,\\
 b\mS     & = & b\pS+\left[(b\pN-b\pS)A
               +(\beta\pN+\beta\pS)(\beta\pN-\beta\pS)^3\right]B,\\
 A_t^-    & = & A_t^+ + (\beta\pN-\beta\pS)^3B,\\
 \beta\mN & = & \beta\pN-(\beta\pN-\beta\pS)AB,\\
 \beta\mS & = & \beta\pS+(\beta\pN-\beta\pS)AB,
\end{eqnarray}
\end{subequations}
with
\begin{eqnarray}
 A & := & b\pN-b\pS+8\omega^+\rh+2A_t^+(\beta\pN-\beta\pS),\\
 \label{B_eqn} B & := & \frac{8\omega^+\rh}{A^2+(\beta\pN-\beta\pS)^4},\\
 \label{A_tpm} A_t^\pm &:= & A_t|\pmN =A_t|\pmS.
\end{eqnarray}
In other words, we are able to express the above boundary values at $\mathcal H^-$ completely
in terms of those on $\mathcal H^+$. These relations are essential for expressing the inner Cauchy horizon potentials entirely in terms of the event horizon potentials, see Sec.~\ref{sec:EP}.

Note that the agreements in \eqref{A_tpm},
\[ A_t|\pN=A_t|\pS,\quad A_t|\mN=A_t|\mS\]
emerge as a consequence of \eqref{BC} and \eqref{RC1}.

In Sec.~\ref{sec:eq} we will derive expressions for angular momentum $J$ and
charge $Q$ which show that
$B$ can be rewritten as
\begin{equation}
B=\frac{8\pi^2\rh}{(\omega^+)^3 [(8\pi J)^2+(4\pi Q^2)^2]},
\end{equation}
i.e. $B$ --- and therefore the quantities in \eqref{minus} --- are
well-defined as long as  $J$ and $Q$ do not both vanish (remember $\omega^+\neq 0, \rh>0$).
 
\section{Ernst potential and electromagnetic potential on the Cauchy
horizon\label{sec:EP}}

From the pseudopotential ${\bf\Omega}$, we now calculate the
potentials $\E$ and $\Phi$ on $\mathcal H^-$. In a first step, we
express $C_1$, $C_2$, and $C_3$ in terms of the event horizon
potentials. 

At the branch points $K_1=\ii\bar z$ and $K_2=-\ii z$, ${\bf\Omega}$ is unique, 
i.e. the values in both $K$-sheets coincide. 
In particular, for $\varrho=0$ (where $K_1=K_2=\zeta$) we have 
\begin{equation}\label{sheet}
 \psi_i^>=\psi_i^<, \quad i=1,2,3,\quad
 \textrm{for}\quad K=\zeta.
\end{equation}
Considering these conditions at $\mathcal H^+$, it follows that (cf.~\eqref{sol2})
\begin{subequations}\label{sheet1}
\begin{align}
 &(\bar\E +2|\Phi|^2)D_1(\zeta)+D_2(\zeta)+\Phi D_3(\zeta)\nonumber\\
  & \quad
  =  (\bar\E +2|\Phi|^2)D_4(\zeta)+D_5(\zeta)+\Phi D_6(\zeta),\\[1.5ex]
 & \E D_1(\zeta)-D_2(\zeta)-\Phi D_3(\zeta)
   =  -\E D_4(\zeta)
   +D_5(\zeta)+\Phi D_6(\zeta),\\[1.5ex]
 &2\bar\Phi D_1(\zeta)+D_3(\zeta) =  2\bar\Phi D_4(\zeta)+D_6(\zeta),
\end{align}
\end{subequations}
where $\E=\E(\zeta)$ and $\Phi=\Phi(\zeta)$ are the potentials taken on
$\mathcal H^+$. 
Using~\eqref{D}, we obtain a linear system of equations for 
$C_1(\zeta)$, $C_2(\zeta)$, $C_3(\zeta)$ with $\zeta\in[-2\rh,2\rh]$.
The corresponding solution reads
\begin{subequations}\label{C}
\begin{eqnarray}\label{C1}
 C_1(\zeta) & = & n
       \left[\bar\E+2\Phi\pN\bar\Phi-2\ii\omega^+(\zeta-2\rh)+\E\pN\right],
       \qquad\\
 C_2(\zeta) & = & (\zeta^2-4\rh^2)^3+n 
   \left[(\E\pN+2|\Phi\pN|^2)(\bar\E+2\Phi\pN\bar\Phi+\E\pN)\right.\nonumber\\
   && \left.+2\ii\omega^+(\zeta-2\rh)\bar\E\,\right],\quad\\
 C_3(\zeta) & = & -2n
      \left[\bar\Phi\pN(\bar\E+2\Phi\pN\bar\Phi+\E\pN) 
       -2\ii\omega^+(\zeta-2\rh)\bar\Phi\,\right],
\end{eqnarray}
\end{subequations}
with
\begin{equation}
 n:=\frac{(\zeta-2\rh)(\zeta+2\rh)^3}{4(\omega^+)^2\ee^{2U}}
    \left[\E+2\bar\Phi\pN\Phi+2\ii\omega^+(\zeta-2\rh)
    -\E\pN-2|\Phi\pN|^2\right].
\end{equation}
Now, we evaluate \eqref{sheet} on $\mathcal H^-$. Similarly to
\eqref{sheet1}, we obtain 
\begin{subequations}\label{sheet2}
\begin{align}
 & (\bar\E +2|\Phi|^2)\tilde D_1(\zeta)+\tilde D_2(\zeta)
   +\Phi \tilde D_3(\zeta)\nonumber\\
 & \quad = (\bar\E +2|\Phi|^2)\tilde D_4(\zeta)+\tilde D_5(\zeta)+\Phi
 \tilde D_6(\zeta),\label{Ga}\\[1.5ex]
 & \E \tilde D_1(\zeta)-\tilde D_2(\zeta)-\Phi \tilde D_3(\zeta) 
   = -\E \tilde D_4(\zeta)+\tilde D_5(\zeta)+\Phi\tilde
 D_6(\zeta),\label{Gb}\\[1.5ex]
 & 2\bar\Phi \tilde D_1(\zeta)+\tilde D_3(\zeta) 
   =  2\bar\Phi \tilde D_4(\zeta)+\tilde D_6(\zeta),\label{Gc}
\end{align}
\end{subequations}
where $\E=\E(\zeta)$ and $\Phi=\Phi(\zeta)$ now denote the potentials on
$\mathcal H^-$. We solve \eqref{sheet2} for the two potentials and get
\begin{subequations}\label{EP}
\begin{eqnarray}
 \bar\Phi(\zeta) & = & \frac{\tilde D_6(\zeta)-\tilde D_3(\zeta)}
                        {2[\tilde D_1(\zeta)-\tilde D_4(\zeta)]},\\[1.5ex]
 \bar\E(\zeta)   & = & \frac{\tilde D_5(\zeta)-\tilde D_2(\zeta)}
                        {\tilde D_1(\zeta)-\tilde D_4(\zeta)}.
\end{eqnarray}
\end{subequations}
Finally, using the expressions for $\tilde D_i$ in terms of $C_i$ and
Eq.~\eqref{C}, we obtain the potentials on $\mathcal H^-$ in terms
of the potentials on $\mathcal H^+$. We arrive at
\begin{subequations}\label{CHpots}
\begin{eqnarray}
 \E^-\!(\theta) & \!\!= & \!\!
 \frac{a_1(\theta) \E^+\!(\pi-\theta)+a_2(\theta)\Phi^+\!(\pi-\theta)
       +a_3(\theta)} 
  {c_1(\theta)\E^+\!(\pi-\theta)+c_2(\theta)\Phi^+\!(\pi-\theta)
  +c_3(\theta)},
  \ \qquad\\[1.5ex]
 \Phi^-\!(\theta) & \!\!= & \!\!
 \frac{b_1(\theta)\E^+\!(\pi-\theta)+b_2(\theta)\Phi^+\!(\pi-\theta)
 +b_3(\theta)} 
  {c_1(\theta)\E^+\!(\pi-\theta)+c_2(\theta)\Phi^+\!(\pi-\theta)
  +c_3(\theta)}, 
\end{eqnarray}
\end{subequations}
in which the inner Cauchy horizon potentials are given with respect to the Boyer-Lindquist-type coordinate $\theta$. As before, the superscripts `$+$' and `$-$' indicate quantities on $\mathcal H^+$ and $\mathcal H^-$, respectively. 
The functions $a_i$, $b_i$, $c_i$, $i=1,2,3$, are given by
\begin{eqnarray*}
 a_1 & = & 16\omega^+\omega^-\rh^2\sin^2\!\theta - 4\ii\omega^+\rh
            F\mN(1+\cos\theta) 
           -4\ii\omega^-\rh F\pN(1-\cos\theta)- F\mN E,\\
 a_2 & = & -2\left[4\ii\omega^+\rh\bar\Phi\mN F\mN(1+\cos\theta)
           +4\ii\omega^-\rh\bar\Phi\pN F\pN(1-\cos\theta)
           +\bar\Phi\pN F\mN E\right],\\
 a_3 & = & -4\ii\omega^+\rh\bar\E\mN F\mN(1+\cos\theta)
           -4\ii\omega^-\rh\bar\E\pN F\pN(1-\cos\theta)-\bar\E\pN F\mN E,\\
 b_1 & = & 4\ii\omega^+\rh\Phi\mN(1+\cos\theta)+4\ii\omega^-\rh\Phi\pN(1-\cos\theta)
           +\Phi\mN E,\\
 b_2 & = & 2\left[8\omega^+\omega^-\rh^2\sin^2\!\theta
           +4\ii\omega^+\rh|\Phi\mN|^2(1+\cos\theta)\right.\\
       &&  \left.+4\ii\omega^-\rh|\Phi\pN|^2(1-\cos\theta)
           +\bar\Phi\pN\Phi\mN E\right],\\
 b_3 & = & 4\ii\omega^+\rh\bar\E\mN\Phi\mN(1+\cos\theta)
           +4\ii\omega^-\rh\bar\E\pN\Phi\pN(1-\cos\theta)
           +\bar\E\pN\Phi\mN E,\\
 c_1 & = & 4\ii\omega^+\rh(1+\cos\theta)+4\ii\omega^-\rh(1-\cos\theta)+E,\\
 c_2 & = & 2\left[4\ii\omega^+\rh\bar\Phi\mN(1+\cos\theta)
           +4\ii\omega^-\rh\bar\Phi\pN(1-\cos\theta)
           +\bar\Phi\pN E\right],\\
 c_3 & = & 16\omega^+\omega^-\rh^2\sin^2\!\theta
           +4\ii\omega^+\rh\bar\E\mN(1+\cos\theta)
           +4\ii\omega^-\rh\bar\E\pN(1-\cos\theta)+\bar\E\pN E,
\end{eqnarray*}
where
\begin{equation*}
 E:=\bar\E\pN-\bar\E\mN+2\Phi\pN(\bar\Phi\pN-\bar\Phi\mN),\quad
 F\pmN:=\bar\E\pmN+2|\Phi\pmN|^2.
\end{equation*}

Note that we have taken only \eqref{Ga} and \eqref{Gc} to obtain
\eqref{EP}. However, using \eqref{CHpots}, we find that \eqref{Gb} is satisfied as well.

\section{A universal equality\label{sec:eq}}

Eqn.~\eqref{ApAm} contains the following black hole quantities:
(i) angular momentum $J$, (ii) electric charge $Q$, and (iii) the two horizon surface areas $A^\pm$. While the expressions for $Q$ and $A^\pm$ are defined unambiguously, the introduction of the angular momentum $J$ requires a bit of explanation. 

The total angular momentum of the spacetime is composed of matter, electromagnetic field and black hole contributions. While clearly the matter part should be excluded for the definition of the local black hole's angular momentum, both a Komar integral and an appropriate electromagnetic event horizon integral must be taken into account, in order to find a measure for which Eqn.~\eqref{ApAm} turns out to be true. A more thorough discussion of this issue is given in \cite{Ansorg}, at the beginning of Sec.~4, and we here adapt the corresponding expression for the local black hole angular momentum $J$ given there.

In terms of the quantities $\hat u$, $\omega$, $A_\varphi$ and $\phi$ we thus obtain (cf. \cite{Ansorg}) 
\begin{subequations}
\begin{eqnarray}
 \label{J}
 J & = & \frac{1}{8\pi}\oint_\Hp(\eta^{i;j}+2\eta^k A_k F^{ij})
       \dd S_{ij}\nonumber\\  
   & = & -\frac{1}{4}\int\limits_0^\pi\hat u\Big[\frac{\hat u}{4}\omega_{,R}
       \sin^2\!\theta
       -A_\varphi(\phi_{,R}-A_\varphi\omega_{,R})\Big]
       \Big|_\Hp \sin\theta\,\dd\theta,\\
 \label{Q}
 Q & = & -\frac{1}{4\pi}\oint_\Hp F^{ij}\dd S_{ij}
    =  -\frac{1}{4}\int\limits_0^\pi
       \hat u(\phi_{,R}-A_\varphi\omega_{,R})\big|_\Hp
       \sin\theta\,\dd\theta,\\
 \label{A}
 A^\pm & = & 2\pi\int\limits_0^\pi\sqrt{\hat\mu\hat u}\big|_\Hpm
         \sin\theta\,\dd\theta = 4\pi\hat u\pmN,
\end{eqnarray}
\end{subequations}
where we have used conditions \eqref{BC} and \eqref{RC}.
As in Sec.~\ref{sec:Intro}, $\eta^i$ is the Killing vector with respect to axisymmetry.

In order to show the validity of Eqn.~\eqref{ApAm}, we express at first 
$J$, $Q$, and $A^\pm$ in terms of the complex potentials $\E$ and $\Phi$. Using \eqref{func},
\eqref{ab2},  and \eqref{phi},
we can perform the integrations in \eqref{J} and \eqref{Q}, i.e. we find
expressions depending only on values on the north and south poles of the
horizons $\Hpm$:
\begin{subequations}
\begin{eqnarray}
 \label{J_eqn}J & = &
 \frac{1}{8(\omega^+)^2}\left[b\pS-b\pN-8\omega^+\rh
   +2A_t^+(\beta\pS-\beta\pN)\right],\\
 \label{Q_eqn}Q & = & \frac{1}{2\omega^+}(\beta\pN-\beta\pS),\\
 A^\pm & = & \pm\frac{32\pi\rh}{\f_{,R}\big|\pmN},
\end{eqnarray}
\end{subequations}
where
\begin{equation}
 \omega^+ = \frac{1}{4}\left[b_{,R}+2(A_t\beta_{,R}-\beta
               A_{t,R})\right]\pN
\end{equation}
arises from \eqref{func}, \eqref{ab2},  and \eqref{BC}.

In order to calculate $\f_{,R}\big|\pmN$, we use the solution of the LP
on $\mathcal A^+$. Evaluation of the conditions in \eqref{sheet},
considered on $\mathcal A^+$, leads us to 
\begin{subequations}\label{sheet3}
\begin{eqnarray}
 (\bar\E +2|\Phi|^2)C_1(\zeta)+C_2(\zeta)+\Phi C_3(\zeta) & = & \psi(\zW),\\
 \E C_1(\zeta)-C_2(\zeta)-\Phi C_3(\zeta) & = & \psi(\zW),\\
 2\bar\Phi C_1(\zeta)+C_3(\zeta) & = & 0.
\end{eqnarray}
\end{subequations}
Summming up the first two of these equations we get
\begin{equation}
 \f(\zeta)=\frac{\psi(\zeta)}{C_1(\zeta)} \quad\textrm{on}\quad \mathcal A^+.
\end{equation}
With the explicit expression \eqref{C1} for $C_1(\zeta)$, we thus obtain both
areas $A^\pm$ in terms of values on the event horizon's north pole:
\begin{subequations}\label{area}
\begin{eqnarray}
 A^+ & = & -\frac{2\pi}{(\omega^+)^2}\,\f_{,\theta\theta}\big|\pN\\
 A^- & = &  -\frac{\pi}{2(\omega^+)^2\,\f_{,\theta\theta}\big|\pN}
            \Big[(\beta\pN-\beta\pS)^4\\
       &&     +\left.\left(b\pS-b\pN-8\omega^+\rh
            +2A_t^+(\beta\pS-\beta\pN)\right)^2\right].\ \qquad
\end{eqnarray}
\end{subequations}
Here, we have used that
\[\f_{,\theta\theta}\big|\pN=\f_{,\theta\theta}\big|\pS\]
and
\[\omega^+ = \frac{1}{4\rh}[b_{,\theta\theta}+2(A_t\beta_{,\theta\theta}
 -\beta A_{t,\theta\theta})]\pN,\]
which can be derived from \eqref{RC}
and regularity conditions, that result from the Ernst equations \eqref{EM_2},
studied at the north and south poles of $\mathcal H^+$.

As both the product $A^+ A^-$  as well as $J$ and $Q$ can be expressed in terms of the same north pole quantities (cf. \eqref{area}, \eqref{J_eqn}, \eqref{Q_eqn}), the validity of the relation \eqref{ApAm} in question can easily be seen.

\section{Discussion\label{sec:Dis}}

We have investigated the interior hyperbolic region 
of axisymmetric and stationary black holes with surrounding matter in Einstein-Maxwell theory.
With the help of the LP for the corresponding Ernst equations,
we have found the explicit relation \eqref{CHpots}
for the complex metric and electromagnetic potentials $\E$ and $\Phi$
on the inner Cauchy horizon $\mathcal H^-$
in terms of those on the event horizon $\mathcal H^+$. 

A discussion of \eqref{CHpots} reveals that with potentials
that are regular on $\mathcal H^+$, the potentials on $\mathcal H^-$ are also regular,
provided that $J$ and $Q$ do not both vanish. In the limit of vanishing $J$ and $Q$,
the potentials $\E^-$ and $\Phi^-$ diverge (cf. the remark at the end of Sec.~\ref{subsec:Hm}).

As an additional result, we have proved a remarkable universal equality for such black
holes. Combining our work with a closely related inequality obtained in \cite{Hennig3},
we arrive at the following.
\begin{Thm}\label{theorem}
	Every regular axisymmetric and stationary Einstein-Maxwell
	black hole with surrounding matter has a regular inner Cauchy horizon if and 
	only if the angular momentum $J$ and charge $Q$ do not both vanish. Then the universal relation
 	$$(8\pi J)^2+(4\pi Q^2)^2 = A^+ A^-$$
 	is satisfied where $A^+$ and $A^-$ denote the areas of event and inner Cauchy horizon
 	respectively.
 	If in addition the black hole is sub-extremal (i.e.~if there exist
 	trapped surfaces in every sufficiently small interior vicinity of the
	event horizon), then the following inequalities hold:
 	$$A^-<\sqrt{(8\pi J)^2+(4\pi Q^2)^2}<A^+.$$
\end{Thm}
Note that in the degenerate limit the above equality becomes identical
with the aforementioned inequalities. As indicated in Sec.~\ref{sec:BC},
the black hole degenerates if the coordinate radius $\rh$ tends to
zero. In this limit the hyperbolic region disappears and the two
horizons $\mathcal H^\pm$ become identical which in turn means
$A^+=A^-$. Then the two formulae in theorem \ref{theorem} yield the
known relation for degenerate axisymmetric and stationary black holes
with surrounding matter in Einstein-Maxwell theory, see \cite{Ansorg}. 

\subsection*{Acknowledgments}
We would like to thank Gernot Neugebauer,
Piotr T. Chru\'sciel and David Petroff for many valuable discussions.
This work was supported by the Deutsche
For\-schungsgemeinschaft (DFG) through the
Collaborative Research Centre SFB/TR7
`Gravitational wave astronomy'.

\end{document}